\documentclass[12pt]{article}
\usepackage[normalem]{ulem}
\usepackage{amsmath}
\usepackage{amssymb}
\usepackage{color}
\usepackage{cite}
\begin{document}
\begin{center}
{\large\bf  {A $3D$ Field-Theoretic Model: Discrete Duality Symmetry}}

\vskip 3.0cm

{\sf R. Kumar$^{(a)}$, R. P. Malik$^{(b,c)}$}\\
$^{(a)}${\textit {Department of Physics, Siksha Bhavana,\\
Visva-Bharati, Santiniketan, Bolpur--731235, West Bengal, India}}\\

\vskip 0.1cm
$^{(b)}${\textit {Department of Physics, Institute of Science,\\ Banaras Hindu University,
Varanasi--221005, Uttar Pradesh (U. P.), India}}\\
\vskip 0.1cm

$^{(c)}${\textit{DST Centre for Interdisciplinary Mathematical Sciences,\\
 Institute of Science, Banaras Hindu University, Varanasi--221005, U. P., India}}\\
 \vskip 0.1cm
 
{\small {\sf {e-mails: rohit.kumar@visva-bharati.ac.in; rpmalik1995@gmail.com}}}

\end{center}

\vskip 2cm

\noindent
{\bf Abstract:}
 We demonstrate the discrete duality symmetry between the Abelian 1-form and 2-form {\it basic} gauge fields in the 
context of a three $(2 + 1)$-dimensional ($3D$) combined system of the field-theoretic model 
of the free Abelian 1-from and 2-form gauge theories within the framework of Becchi-Rouet-Stora-Tyutin 
(BRST) formalism. The classical gauge-fixed Lagrangian density of {\it this} theory is generalized to its quantum 
counterpart as the BRST and co-BRST invariant Lagrangian density. We show clearly the existence of the off-shell 
nilpotent (co-)BRST symmetry transformations and establish their intimate connection through a set of underlying 
discrete duality symmetry transformations in our $3D$ BRST-quantized theory. We provide the mathematical {\it basis} 
for the existence of the discrete duality symmetry transformations in our theory through the Hodge duality 
operator (that is defined on the $3D$ flat Minkowskian spacetime manifold). We briefly mention a bosonic symmetry 
transformation which is constructed from the anticommutator of the above off-shell nilpotent (co-)BRST symmetry 
transformations. We lay emphasis on the algebraic structures of the existing continuous and discrete duality 
symmetry transformations for our $3D$ BRST-quantized theory (where they are treated as operators). We also comment on the 
appearance of a pseudo-scalar field (with negative kinetic term). This field happens to be one of the possible candidates 
for the phantom field of the cosmological models.


\newpage


\section{Introduction}
\label{sec1}

The study of the higher $p$-form ($ p = 2, 3, 4,...$) gauge theories\footnote{The most successful theory of 
THEP is the standard model of particle physics (SMPP)  which is based on the {\it interacting} non-Abelian 
1-form (i.e. $p = 1$) gauge theory where there is a stunning degree of agreements between {\it this} theory and experiments. 
However, the conclusive experimental evidence of the masses of the neutrinos has compelled the modern-day 
practitioners of THEP to go beyond the domain of the validity of SMPP. The ideas behind the (super)string 
theories are very promising in this direction.} has become important and essential due to the appearance of their 
{\it basic} fields in the quantum excitations of (super)strings which are the forefront areas of research 
activities in the realm of theoretical high energy physics (THEP). The modern developments in the domain of 
(super)string theories (see, e.g.~\cite{RPM1,RPM2,RPM3,RPM4,RPM5} and references therein) have brought {\it together}, 
in an unprecedented manner,  the active researchers in THEP and pure mathematics on a single intellectual platform where {\it both} 
kinds of researchers have benefited from the knowledge and creativity of each-other. In other words, 
there has been a notable convergence of ideas from the research activities in the realms of THEP and pure mathematics. 
Such kinds of confluence of ideas have been observed in our studies of the massless and St{\" u}ckelberg-modified 
massive  Abelian $p$-form (i.e. $p = 1, 2, 3 $) gauge theories (see, e.g.~\cite{RPM6,RPM7,RPM8,RPM9} and references therein) 
 within the framework of Becchi-Rouet-Stora-Tyutin (BRST) formalism~\cite{RPM10,RPM11,RPM12,RPM13} where it has been  established that, 
in the $ D = 2 p$ (i.e. $D = 2, 4, 6 $) dimensions of spacetime, {\it these} theories turn out to be the 
field-theoretic examples for Hodge theory. The discrete and continuous symmetry transformations 
(and corresponding conserved  charges) of these theories have been able to provide the physical realizations of 
the de Rham cohomological operators of differential geometry (see, e.g.~\cite{RPM14,RPM15,RPM16,RPM17,RPM18}) at the {\it algebraic} level. 
In other words, in our research activities, it has been established that there is  a convergence
of ideas from the physical aspects of the BRST formalism and the mathematical ingredients related to
the cohomological operators of differential geometry. It is crystal clear, from our present discussions, 
that our studies~\cite{RPM6,RPM7,RPM8,RPM9} on the above BRST-quantized Abelian theories have been confined {\it only} to the 
{\it even} dimensions (i.e. $D = 2, 4, 6$) of the flat Minkowskian  spacetime.

The central purpose of our present investigation is to focus on an {\it odd} dimensional field-theoretic example  
where we consider the {\it combined} system of a three $(2 + 1)$-dimensional ($3D$) free Abelian 1-form and 2-form 
gauge theories (within the framework of BRST formalism) to demonstrate  that there is a set of discrete duality 
symmetries in {\it this} theory which connects the {\it basic} Abelian 1-form and 2-form gauge fields in a 
meaningful manner (i) at the level of the {\it classically} gauge-fixed Lagrangian density [cf. Eq.~\eqref{12} below],
and (ii) at the quantum level which is described by the (co-)BRST invariant Lagrangian density [cf. Eq.~\eqref{21} below]. 
This observation plays a crucial role in providing  the physical realizations of the algebraic structures of the 
cohomological operators of differential geometry. To be precise, {\it these} duality symmetry transformations connect
the off-shell nilpotent BRST and co-BRST symmetry transformations that are respected by the BRST-invariant Lagrangian 
density of our theory [cf. Eq.~\eqref{21} below]. The {\it latter} also respects a set of bosonic symmetry transformations 
that is obtained from the anticommutator of the off-shell nilpotent BRST and co-BRST symmetry transformations. 
One of the highlights of our present endeavor is the observation that the operator forms of the nilpotent 
(co-)BRST and non-nilpotent bosonic symmetry transformations obey an algebra that is reminiscent of the algebra 
that is satisfied by the de Rham cohomological operators of differential geometry.

The existence of the discrete duality symmetry transformations  in our $3D$ theory is very important 
as they provide (i) a beautiful {\it algebraic} relationship between the off-shell nilpotent (co-)BRST 
symmetry transformations (cf. Sec.~\ref{sec4}), and (ii) a {\it direct} connection between the fermionic 
(i.e. nilpotent) (co-)BRST symmetry transformations of our $3D$ theory (cf.~\ref{secA}).  In our recent 
couple of papers~\cite{RPM19,RPM20}, we have not been able to obtain such kinds of relationships due to the fact that 
we were {\it unaware} of the existence of such  beautiful types of discrete duality symmetry transformations. 
As a consequence, we could not provide the physical realization of the Hodge duality operator 
(in our earlier works~\cite{RPM19,RPM20}) which plays a key role in the algebraic structures
that are satisfied by the de Rham cohomological operators of differential geometry. In our present endeavor, 
we have given a great deal of importance to the discrete duality symmetry transformations because they provide 
(i) a deep connection between the Abelian 1-form and 2-form basic gauge fields\footnote{We would like to lay emphasis on the 
{\it physical} importance of this duality between the {\it basic} Abelian  1-form and 2-form gauge fields because this 
observation [cf. Eqs.~\eqref{16},\eqref{17}]  is
at the level of the gauge-fixed Lagrangian density [cf. Eq.~\eqref{12}] which is completely  different from the {\it usual}
Hodge duality between the 1-form and 2-from that exist {\it mathematically}  on  a given $3D$ compact spacetime manifold. }
[cf. Eq.~\eqref{20}], and
(ii) the physical realization of the Hodge duality $*$ operator~\cite{RPM14,RPM15,RPM16,RPM17,RPM18} in the relationship: 
$\delta = \pm\, *\, d\, * $ between the (co-)exterior derivatives [i.e. $(\delta)d$].

One of the key upshots of our present endeavor is the observation that there is existence of the massless 
(i.e. $\Box \widetilde \phi = 0, \; \Box  \phi = 0 $) (pseudo-)scalar fields [i.e. $(\widetilde \phi)\phi $] 
in our theory which carry the kinetic terms that are endowed with the {\it opposite} signs. However, despite 
this obvious distinct difference, they obey the {\it massless} Klein-Gordon equations of motion: 
$\Box \widetilde \phi = 0, \; \Box  \phi = 0 $. It is interesting to point that the fields with 
{\it negative} kinetic terms have become quite popular in the realm of the cosmological models of the Universe. 
In our theory, on the symmetry grounds {\it alone},  we observe that the pseudo-scalar field ($\widetilde \phi $) 
carries the {\it negative} kinetic term. Such kinds of fields are a set of  possible candidates for 
(i) the ``phantom'' and/or ``ghost'' fields in the realm of the cyclic, bouncing and self-accelerated 
cosmological models of the Universe~\cite{RPM21,RPM22,RPM23,RPM24,RPM25}, and (ii) the dark energy 
(see, e.g.~\cite{RPM26,RPM27} and references therein) 
because these fields automatically lead to the creation of the negative pressure in the theory (which happens 
to be one of the  characteristic features of dark energy). On the contrary, the pure scalar field ($ \phi $), with the 
{\it positive} kinetic term, is a {\it normal} field which does {\it not} lead to the existence of the negative 
pressure in the theory. Hence, the existence of the pseudo-scalar field (in our theory)  is 
very interesting in the current scenario of research activities that are related to (i) the realms of 
the cosmological models of the Universe, and (ii) the search for the possible dark energy candidate(s).

The theoretical materials of our present investigation are organized as follows. In Sec.~\ref{sec2}, we provide a brief 
synopsis of the first-class constraints and ensuing classical gauge symmetry transformations for the combined system of 
the $3D$ free Abelian 1-form and 2-form gauge theories. In this section, we also discuss the gauge-fixed classical 
Lagrangian density and the continuous as well as discrete duality symmetry transformations associated with it. 
Our Sec.~\ref{sec3} deals with the elevation of the gauge-fixed {\it classical} Lagrangian density to its 
quantum counterpart as the (co-)BRST invariant Lagrangian density (where we show explicitly the existence of 
the nilpotent (co-)BRST symmetry transformations). The subject matter of our Sec.~\ref{sec4} is connected with the 
derivations of (i) a bosonic symmetry transformation which is constructed  from the anticommutator of the 
(co-)BRST transformations, and (ii) the algebraic structures amongst the symmetry operators. The {\it latter}  
establish an intimate connection between the BRST and co-BRST transformations through the application of the 
discrete duality symmetry transformations. Finally, in our Sec.~\ref{sec5}, we summarize our key results and point out 
the future perspective of our present investigation.

In our~\ref{secA}, we establish a {\it direct} connection between the off-shell nilpotent BRST and co-BRST symmetry transformations by 
exploiting the theoretical beauty and strength of the discrete duality symmetry transformations of our $3D$ theory. \\


\section{Preliminaries: Constraint Analysis, (Dual-)Gauge and Discrete Duality Symmetry Transformations}
\label{sec2}


We begin with the $3D$ field-theoretic model for the combined system of the free Abelian 1-form (i.e. $A^{(1)} = A_\mu\, dx^\mu$) 
and 2-form [i.e. $B^{(2)} = \frac{1}{2!}\, B_{\mu\nu} (dx^\mu \wedge dx^\nu)$] gauge theories which is described by the following 
standard Lagrangian density\footnote{Our three $(2 + 1)$-dimensional ($3D$) Minkowskian spacetime is characterized by the flat 
metric tensor $\eta_{\mu\nu} = \text{diag} \,(+1, -1, -1)$ so that the dot product between the two non-null Lorentz vectors 
$U_\mu$ and $V_\mu$ is defined as: $U \cdot V = \eta_{\mu\nu}\,U^\mu\,V^\nu \equiv U_0 \,V_0 - U_i\, V_i$ where the Greek indices 
$\mu, \nu, \sigma, ... = 0, 1, 2$ stand for the time and space directions and the Latin indices $i, j, k, ... = 1, 2$ denote the 
space directions only. The $3D$ Levi-Civita tensor $\varepsilon_{\mu\nu\sigma}$ is chosen such that 
$\varepsilon_{012} = \varepsilon^{012} = +1$ and the standard relationships: 
$\varepsilon^{\mu\nu\sigma}\,\varepsilon_{\mu\nu\sigma} = +3!$, 
$\varepsilon^{\mu\nu\sigma}\,\varepsilon_{\mu\nu\eta} = +2!\, \delta^\sigma_\eta$,  
$\varepsilon^{\mu\nu\sigma}\,\varepsilon_{\mu\eta\kappa} 
= +1!\, (\delta^\nu_\eta\, \delta^\sigma_\kappa - \delta^\nu_\kappa\, \delta^\sigma_\eta)$ 
are satisfied. We also adopt the convention of the derivative w.r.t. the antisymmetric tensor field as: 
$\big (\partial B_{\mu\nu}/ \partial B_{\sigma \kappa}\big ) 
= \frac{1}{2!} \, (\delta^\sigma_\mu\, \delta^\kappa_\nu - \delta^\sigma_\nu\, \delta^\kappa_\mu)$, etc.}
\begin{eqnarray}\label{1}
{\cal L}_{(0)} = - \dfrac{1}{4}\, F^{\mu\nu}\, F_{\mu\nu} 
+ \dfrac{1}{12}\, H^{\mu\nu\sigma}\, H_{\mu\nu\sigma} \equiv   
- \dfrac{1}{2}\, \Big(\varepsilon^{\mu\nu\sigma} \,\partial_\nu A_\sigma \Big)^2  
+ \dfrac{1}{2}\,\Big(\dfrac{1}{2}\varepsilon^{\mu\nu\sigma} \,\partial_\mu B_{\nu\sigma} \Big)^2,
\end{eqnarray}
where the field-strength tensors: $F_{\mu\nu} = \partial_\mu A_\nu - \partial_\nu A_\mu$ and 
$H_{\mu\nu\sigma} = \partial_\mu B_{\nu\sigma} +  \partial_\nu B_{\sigma\mu} +  \partial_\sigma B_{\mu\nu}$ 
are derived from the 2-form $F^{(2)} = d\, A^{(1)} = \frac{1}{2!}\, F_{\mu\nu} \big(dx^\mu \wedge dx^\nu \big)$
and 3-form $H^{(3)} = d\, B^{(2)} = \frac{1}{3!}\, H_{\mu\nu\sigma}\,\big(dx^\mu \wedge dx^\nu \wedge dx^\sigma \big)$, respectively. 
Here the symbol $d = \partial_\mu \,dx^\mu$ [with $d^2 = \frac{1}{2!}\,\big(\partial_\mu \, \partial_\nu - \partial_\nu \, \partial_\mu \big) 
\big(dx^\mu \wedge dx^\nu \big)= 0$] denotes the exterior derivative of differential geometry 
(see, e.g.~\cite{RPM14,RPM15,RPM16,RPM17,RPM18}) and the above 1-form (i.e. $ A^{(1)}$) and 2-form (i.e. $B^{(2)}$) 
define the Lorentz vector gauge field $A_\mu$ and the second-rank antisymmetric (i.e. $B_{\mu\nu} =-\, B_{\nu\mu}$) 
tensor gauge field $B_{\mu\nu}$, respectively. It is the specific feature of the three $(2 + 1)$-dimensional $(3D)$ 
Minkowskian spacetime that the kinetic terms for the Abelian 1-form gauge field $A_\mu$ and the antisymmetric tensor 
gauge field $B_{\mu\nu}$ can be expressed in terms of the $3D$ Levi-Civita tensor and the derivative on {\it them} 
as we have expressed {\it these} terms in our equation~\eqref{1}. One of the key reasons behind this observation is 
the fact that, in the $3D$ Minkowskian spacetime, we have: $\frac{1}{12} \,H^{\mu\nu\sigma}\,H_{\mu\nu\sigma} 
= \frac{1}{2}\, H^{012}\, H_{012} \equiv \frac{1}{2}\, (H_{012})^2$.  The component $H_{012}$, however, can be written 
in its covariant form as: $H_{012} = \frac{1}{2}\,\varepsilon^{\mu\nu\sigma}\,\partial_\mu B_{\nu\sigma}$  which has 
been taken into account in our equation \eqref{1} to express the kinetic term for the Abelian 2-form gauge field. 
Similarly, it is straightforward to check that the $3D$ kinetic term for the Abelian 1-form field is: 
$- \frac{1}{4}\, F^{\mu\nu}\, F_{\mu\nu} = - \frac{1}{2}\, \big(\varepsilon^{\mu\nu\sigma} \partial_\nu A_\sigma\big)^2$. 
For our combined system of the $3D$ free Abelian 1-form and 2-form gauge theories, the infinitesimal, local and continuous 
gauge symmetry transformations $(\delta_g)$ are well-known. These transformations for our $3D$ system are
\begin{eqnarray}\label{2}
\delta_g B_{\mu\nu} = -(\partial_\mu \Lambda_\nu - \partial_\mu \Lambda_\nu), \qquad \qquad \delta_g A_\mu = \partial_\mu \Lambda, 
\end{eqnarray}
where the Lorentz vector $\Lambda_\mu(x)$ and the Lorentz scalar $\Lambda (x)$ are the infinitesimal and local 
gauge symmetry transformation parameters. The above infinitesimal gauge symmetry transformations~\eqref{2} 
are generated by the first-class constraints that exist on our theory.

To determine the first-class constraints, the first-step is to derive the canonical 
conjugate momenta \big(i.e. $\Pi^{\mu\nu}_{(B)},\; \Pi^{\mu}_{(A)}$\big)  w.r.t. the basic gauge 
fields $B_{\mu\nu}$ and $A_\mu$, respectively. For the $3D$ version of the Lagrangian density \eqref{1}, these momenta are:
\begin{eqnarray}\label{3}
&&\Pi^{\mu\nu}_{(B)} = \dfrac{\partial {\cal L}_{(0)}}{\partial (\partial_0 B_{\mu\nu})} 
\equiv \dfrac{1}{2}\, \varepsilon^{0\mu\nu}\, \Big(\dfrac{1}{2}\,\varepsilon_{\sigma\eta\kappa}\, \partial^\sigma B^{\eta\kappa} \Big) = \dfrac{1}{2}\,  \varepsilon^{0\mu\nu}\,H_{012} \;\Longrightarrow \; \Pi^{0i}_{(B)} = \dfrac{1}{2}\,  \varepsilon^{00i}\,H_{012} \approx 0, \nonumber\\
 &&\Pi^\mu_{(A)} = \dfrac{\partial {\cal L}_{(0)}}{\partial (\partial_0 A_\mu)} = - \varepsilon^{0\mu\sigma}\, \Big(\varepsilon_{\sigma\nu\kappa} \,\partial^\nu A^\kappa \Big) \equiv  -F^{0\mu} \quad \Longrightarrow \quad \Pi^0_{(A)} = - F^{00} \approx 0. 
\end{eqnarray}
The above equation demonstrates that we have  two sets of {\it primary constraints} 
(i.e. $\Pi^{0i}_{(B)} \approx 0, \; \Pi^0_{(A)} \approx 0$) on our theory where the 
symbol $\approx 0$ stands for {\it weakly} zero in Dirac's notation (see, e.g.~\cite{RPM28,RPM29} for details). 
The requirements of the time-evolution invariance of these primary constraints lead to the 
determination of the secondary constraints. The following Euler-Lagrange (EL) equations of motion (EoMs)
\begin{eqnarray}\label{4}
\partial_\mu \Big(\dfrac{1}{2}\,\varepsilon^{\mu\nu\sigma}\,H_{012}\Big) = 0, \quad\qquad \partial_\mu F^{\mu\nu} =0,
\end{eqnarray}
would lead to the time-evolution invariance of the primary constraints for the choices: 
$\nu = 0,\; \sigma = i$ and $\nu = 0$ in the above equations, respectively. These can be written as
\begin{eqnarray}\label{5}
\partial_0 \Big(\dfrac{1}{2}\,\varepsilon^{00i}\,H_{012}\Big) 
+ \partial_j \Big(\dfrac{1}{2}\,\varepsilon^{j0i}\,H_{012}\Big) & =& 0 \; \Longrightarrow \; \partial_0 \Pi^{0i}_{(B)} 
= \partial_j \Pi^{ji}_{(B)}  \approx 0, \nonumber\\ 
 \partial_0 F^{00} + \partial_i F^{i0}  &=& 0 \; \Longrightarrow  \; \partial_0 \Pi^{0}_{(A)} = \partial_i \Pi^{i}_{(A)} \approx 0,\qquad
\end{eqnarray}
where $\Pi^{ij}_{(B)} = \frac{1}{2}\, \varepsilon^{0ij}\, H_{012}$ and $\Pi^i_{(A)} = -F^{0i}$  are the 
space components of the conjugate momenta that have been defined in our equation~\eqref{3}. It is obvious 
that we have {\it four} sets of constraints on our theory out of which two sets 
(i.e. $\Pi^{0i}_{(B)} \approx 0, \; \Pi^{0}_{(A)} \approx 0$) are the primary constraints and the 
{\it rest} of the two (i.e. $\partial_j \Pi^{ji}_{(B)} \approx 0,\; \partial_i \Pi^i_{(A)} \approx 0$) 
are the secondary constraints~\cite{RPM28,RPM29}. Since all these constraints are expressed in terms of the components 
of the conjugate momenta~\eqref{3}, {\it all} of them commute among themselves. Hence, they belong to the 
first-class variety according to Dirac's prescription for the classification scheme of constraints~\cite{RPM28,RPM29}. 
These first-class constraints generate the infinitesimal gauge symmetry transformations~\eqref{2} 
which we explain in the following paragraph.

According to Noether's  theorem, the existence of the continuous symmetry transformations 
{\it always} leads to the derivations of the Noether conserved currents and corresponding charges. 
Due to the infinitesimal gauge symmetry transformations  in our equation~\eqref{2}, we find the following 
expression for the Noether {\it gauge} current $J^\mu_{(G)}$ for our $3D$ field-theoretic
{\it combined} system of the free Abelian 1-form and 2-form theory
\begin{eqnarray}\label{6}
J^\mu_{(G)} =  - F^{\mu\nu}\,\big(\partial_\nu \Lambda \big) 
 -  \dfrac{1}{2} \varepsilon^{\mu\nu\sigma} H_{012}\, \big(\partial_\nu \Lambda_\sigma - \partial_\sigma \Lambda_\nu\, \big),
\end{eqnarray}
which is conserved (i.e. $\partial_\mu J^\mu_{(G)} = 0$) due the EL-EoMs in \eqref{4} and the antisymmetric 
properties of the tensors: $F^{\mu\nu}$ and $\varepsilon^{\mu\nu\sigma}$. This conserved Noether gauge 
current [cf. Eq.~\eqref{6}]  leads to the definition of the conserved Noether charge 
$Q_G = \int d^2x\, J^0_{(G)}$ as follows:
\begin{eqnarray}\label{7}
Q_G = \int d^2x \Big[ - F^{0\nu} \big(\partial_\nu \Lambda \big) 
- \dfrac{1}{2} \varepsilon^{0\nu\sigma} H_{012} \big(\partial_\nu \Lambda_\sigma - \partial_\sigma \Lambda_\nu\big)\Big].
\end{eqnarray}
Since we have constraints (i.e. $- F^{00} \approx 0, \; 
\frac{1}{2}\, \varepsilon^{00i}\, H_{012} \approx 0$) on our theory, we have to be careful in expressing 
the r.h.s. of the above charge $Q_G$ so that these constraints are not  {\it strongly} set equal to zero. 
In other words, we have the following explicit expression:
\begin{eqnarray}\label{8}
 Q_G &=& \int d^2x \Big[ - F^{00} \big(\partial_0 \Lambda \big)  - F^{0i} \big(\partial_i \Lambda \big) 
- \dfrac{1}{2} \varepsilon^{00i} H_{012} 
\big(\partial_0  \Lambda_i  - \partial_i  \Lambda_0 \big) \nonumber\\
&-& \dfrac{1}{2} \varepsilon^{0ij} H_{012}\big(\partial_i \Lambda_j - \partial_j \Lambda_i \big) \Big]. 
\end{eqnarray}
Taking into account the primary constraints and the space components of the momenta that are present in
equation \eqref{3}, we recast the above charge in the following form:
\begin{eqnarray}\label{9}
Q_G = \int d^2x \Big[\Pi^0_{(A)}\,\big(\partial_0 \,\Lambda \big)  + \Pi^i_{(A)} \big(\partial_i  \Lambda \big) 
- \Pi^{0i}_{(B)}\big(\partial_0 \Lambda_i 
- \partial_i \Lambda_0 \big) 
- \Pi^{ij}_{(B)}\big(\partial_i \Lambda_j - \partial_j \Lambda_i \big) \Big]. 
\end{eqnarray}
Using the following non-zero equal-time commutators\footnote{At the classical level, we should use 
the Poisson-brackets instead of the canonical commutators. However, since we shall be discussing the 
BRST-quantized version of our present classical $3D$ theory, we shall stick with the canonical 
commutators which will be useful for our discussions at the {\it quantum} level.}
\begin{eqnarray}\label{10}
\big[B_{0i}(\vec x, t), \, \Pi^{0j}_{(B)}(\vec y, t)\big] &=& i\,\delta^j_i\, \delta^{(2)}(\vec x- \vec y), \nonumber\\
 \big[B_{ij}(\vec x, t), \, \Pi^{kl}_{(B)}(\vec y, t)\big] &=& \dfrac{i}{2} \big(\delta^k_i \delta^l_j 
- \delta^k_j \delta^l_i \big) \delta^{(2)}(\vec x- \vec y), \nonumber\\
\big[A_0(\vec x, t), \, \Pi^0_{(A)}(\vec y, t)\big] &=& i\, \delta^{(2)}(\vec x- \vec y), \nonumber\\
\big[A_i(\vec x, t), \, \Pi^j_{(A)}(\vec y, t)\big] &=& i\, \delta^j_i\, \delta^{(2)}(\vec x- \vec y), 
\end{eqnarray}
(in the natural units where we have: $\hbar = c = 1$), it is straightforward to check that we obtain the 
infinitesimal gauge symmetry transformations~\eqref{2} from the conserved Noether charge
$Q_G$ [cf. Eq.~\eqref{9}] if we use the standard relationship between the continuous symmetry 
transformation $(\delta_g)$ and its generator $Q_G$ as: $\delta_g \Phi(\vec x, t) = - i \,\big[\Phi(\vec x, t),\; Q_G\big]$
where the generic field $\Phi(\vec x, t) \equiv A_0(\vec x, t),\, A_i(\vec x, t), \, B_{0i}(\vec x, t),\, B_{ij} (\vec x, t)$. 
The above expression for the conserved Noether charge~\eqref{9} can be re-written in terms of the 
first-class constraints [cf. Eqs.~\eqref{3}, \eqref{5}] as the generator $G$ of the infinitesimal gauge 
symmetry transformations \eqref{2} by exploiting the theoretical strength of the Gauss divergence theorem as:
\begin{eqnarray}\label{11}
Q_G \to G &=& \int d^2x \Big[\Pi^0_{(A)} \big(\partial_0 \Lambda \big)  
-  \big(\partial_i \Pi^i_{(A)} \big)  \Lambda -  \Pi^{0i}_{(B)} \big(\partial_0 \Lambda_i - \partial_i \Lambda_0 \big) \nonumber\\
&+&  \big(\partial_i \Pi^{ij}_{(B)} \big) \Lambda_j +  \big(\partial_j \Pi^{ji}_{(B)} \big)  \Lambda_i \Big], 
\end{eqnarray}
which matches with the general expression for the generator $G$, in terms of the first-class constraints, 
that has been written in a nice paper (see, e.g.~\cite{RPM30} for details).

For the quantization of our combined system of the free Abelian 1-form and 2-form
gauge theories, we have to incorporate the gauge-fixing terms [i.e. ${\cal L}_{(gf)}$] into the starting Lagrangian
density [cf. Eq.~\eqref{1}]. The ensuing gauge-fixed Lagrangian density is:
\begin{eqnarray}\label{12}
{\cal L}_{(0)} + {\cal L}_{(gf)} =  - \dfrac{1}{2} \Big(\varepsilon^{\mu\nu\sigma} \,\partial_\nu A_\sigma \Big)^2 
- \dfrac{1}{2} \big(\partial \cdot A \big)^2 
+ \dfrac{1}{2}\Big(\dfrac{1}{2}\varepsilon^{\mu\nu\sigma} \,\partial_\mu B_{\nu\sigma} \Big)^2 
+ \dfrac{1}{2} \big(\partial^\nu B_{\nu\mu} \big)^2. 
\end{eqnarray}
The gauge-fixing terms in the above owe their origin to the co-exterior derivative $\delta = \pm\, *\, d\, *$ 
of the differential geometry~\cite{RPM14,RPM15,RPM16,RPM17,RPM18}. For instance, it can be readily checked that: 
$\delta \,A^{(1)} = \pm \,*\, d\, * \big(A_\mu\, dx^\mu \big) \equiv \pm \big(\partial \cdot A \big)$ 
and $\delta \,B^{(2)} = \pm\, * \,d\, * \big[\frac{1}{2!}\,B_{\mu\nu}\, \big(dx^\mu \wedge dx^\nu\big) \big] 
\equiv \mp \big(\partial^\nu B_{\nu\mu} \big) dx^\mu$  where $*$ is the Hodge duality operator on our 
chosen $3D$ Minkowskian spacetime manifold. Under the infinitesimal local gauge symmetry 
transformations~\eqref{2}, we obtain the following:
\begin{eqnarray}\label{13}
\delta_g \big[{\cal L}_{(0)} + {\cal L}_{(gf)} \big] 
= \partial_\mu \Big[\big(\partial_\nu B^{\nu\mu}\big)\big(\partial \cdot \Lambda \bigl) \Big] 
- \big(\partial \cdot A \big) \,\Box \Lambda - \big(\partial_\nu B^{\nu\mu} \big)\, \Box \Lambda_\mu. 
\end{eqnarray} 
The above equation demonstrates that if we impose the conditions:  $\Box \Lambda = 0$ and  $\Box \Lambda_\mu = 0$ 
on the Lorentz scalar and vector gauge transformation parameters, we have the gauge invariance in the theory because 
the gauge-fixed Lagrangian density~\eqref{12} transforms to a total spacetime derivative (thereby rendering the 
corresponding action integral invariant). We have another infinitesimal, local and continuous symmetry in the 
theory which we call as the dual-gauge symmetry transformations $(\delta_{dg})$, under which, we have the following 
transformations for the Abelian 1-form and 2-form {\it basic} gauge fields, namely;
\begin{eqnarray}\label{14}
\delta_{dg} A_\mu = - \,\varepsilon_{\mu\nu\sigma}\partial^\nu \Sigma^\sigma, \qquad  \qquad
\delta_{dg} B_{\mu\nu} = + \,\varepsilon_{\mu\nu\sigma} \partial^\sigma \Sigma,
\end{eqnarray}
where the Lorentz axial-vector $\Sigma_\mu(x)$ and pseudo-scalar $\Sigma(x)$ are the infinitesimal 
local dual-gauge symmetry transformation parameters (so that the parity symmetry could be maintained 
at every stage of the above transformations). Under the above infinitesimal dual-gauge symmetry transformations, 
the gauge-fixed Lagrangian density transforms to:
\begin{eqnarray}\label{15}
\delta_{dg} \big[{\cal L}_{(0)} + {\cal L}_{(gf)} \big] 
= \partial_\mu \Big[\varepsilon^{\mu\nu\sigma} \,\partial_\nu A_\sigma \big(\partial \cdot \Sigma\big) \Big] 
 + \frac{1}{2}\, \Big(\varepsilon^{\mu\nu\sigma} \, \partial_\mu B_{\nu\sigma}\Big) \Box \Sigma 
- \Big(\varepsilon^{\mu\nu\sigma} \, \partial_\nu A_\sigma \Big) \Box \Sigma_\mu. 
\end{eqnarray} 
The above equation demonstrates that if we choose the infinitesimal dual-gauge transformation parameters such 
that the restrictions:  $\Box \Sigma_\mu = 0$ and  $\Box \Sigma = 0$ are satisfied  {\it together}, we have a 
{\it perfect} dual-gauge symmetry invariance in the theory because the gauge-fixed Lagrangian density~\eqref{12} transforms to a 
total spacetime derivative (under the above conditions). To be precise, we note that, under the above
restrictions on the transformation parameters, the action integral (corresponding to the gauge-fixed Lagrangian density~\eqref{12})
remains invariant because of the Gauss divergence theorem due to which {\it all} the physical fields as well as the  transformation parameters 
of our theory vanish off as $x \to \pm\, \infty $. In addition to the infinitesimal 
(i) gauge symmetry transformations~\eqref{2}, and (ii) dual-gauge symmetry transformations~\eqref{14} 
(with similar kinds of restrictions on the transformation parameters: 
$\Box \Lambda = 0,\;  \Box \Lambda_\mu = 0, \; \Box \Sigma = 0, \; \Box \Sigma_\mu = 0$), we have the following 
{\it perfect} discrete duality symmetry transformations\footnote{Mathematical origin for the existence of the 
{\it perfect} duality symmetry transformations~\eqref{16} for the $3D$ {\it combined} system of the free Abelian 1-form 
and 2-form gauge theories, seems to be hidden in the Hodge duality $*$ operation on the Abelian 2-form and 1-form, 
respectively. For instance, we note that: 
$*\, B^{(2)} = \big [ \frac{1}{2!} \, \varepsilon_{\mu\nu\sigma}\, B^{\nu\sigma} \big ]\, dx^\mu \sim A_\mu \, dx^\mu$ 
which demonstrate that, for our $3D$ theory, the Hodge dual of 2-form is the 1-form. In a similar manner, 
we observe that the Hodge dual of the 1-form: $*\, A^{(1)} = \frac{1}{2!}\, \big [\varepsilon_{\mu\nu\sigma} \, A^\sigma \big ]\,
(dx^\mu \wedge dx^\nu) \sim \frac{1}{2!} \, B_{\mu\nu}\, (dx^\mu \wedge dx^\nu)$ is the 2-form.
These observations are the reasons behind the {\it numeral factors} that appear in the discrete duality 
transformations~\eqref{16} modulo the overall factors of $\pm\, i$.}: 
\begin{eqnarray}\label{16}
A_\mu \longrightarrow \,\mp\, \dfrac{i}{2}\, \varepsilon_{\mu\nu\sigma}\, B^{\nu\sigma}, 
\quad \qquad B_{\mu\nu} \longrightarrow \,\pm\, i\,\varepsilon_{\mu\nu\sigma}\, A^\sigma.
\end{eqnarray}
This is due to the fact that we note the following
\begin{eqnarray}\label{17}
 -\, \dfrac{1}{2}\, \big(\partial \cdot A \big)^2 \longleftrightarrow  
+ \,\dfrac{1}{2}\,\Big(\dfrac{1}{2}\varepsilon^{\mu\nu\sigma} \,\partial_\mu B_{\nu\sigma} \Big)^2, \qquad
+ \,\dfrac{1}{2}\, \big(\partial^\nu B_{\nu\mu} \big)^2  \longleftrightarrow 
 - \,\dfrac{1}{2}\, \Big(\varepsilon^{\mu\nu\sigma} \,\partial_\nu A_\sigma \Big)^2,
\end{eqnarray}
where (i) the gauge-fixing term of the Abelian 1-form field interchanges with the kinetic term of the 
Abelian 2-form field, and (ii) the gauge-fixing term of the Abelian 2-form field interchanges with the 
kinetic term of the Abelian 1-form field. This observation is very crucial as it will play a key role at
the {\it quantum} level when we shall discuss various kinds of continuous and discrete symmetries of our 
$3D$ field-theoretic system (as the BRST-quantized theory).

We conclude this section with the following remarks. First, under the gauge symmetry transformations~\eqref{2}, 
the kinetic terms (owing their origin to the exterior derivative of differential geometry) remain invariant. 
Second, the gauge-fixing terms (owing their existence to the co-exterior derivative of differential geometry) 
are found to be invariant under the dual-gauge symmetry transformations. Finally, similar kinds of {\it outside} 
restrictions: $\Box \Sigma = 0, \; \Box \Sigma_\mu = 0,\; \Box \Lambda = 0,\;  \Box \Lambda_\mu = 0$ on the 
(dual-)gauge symmetry transformation parameters at the {\it classical} level will be taken care of by the 
(anti-)ghost fields at the {\it quantum} level within the framework of BRST formalism (without any
kinds of  outside restrictions).


\section{BRST and co-BRST Invariant Lagrangian Density}
\label{sec3}


The gauge-fixed Lagrangian density~\eqref{12} for our $3D$ combined system of the free
Abelian 1-form and 2-form gauge theories can be {\it linearized}, in its most general form in two different ways
by invoking the auxiliary fields, as follows
\begin{eqnarray}\label{18}
{\cal L}^{(1)}_{({\cal B}, B)} &=& \dfrac{1}{2}\, {\cal B}^\mu {\cal B}_\mu 
- {\cal B}^\mu \Big(\varepsilon_{\mu\nu\sigma} \partial^\nu A^\sigma 
- \dfrac{1}{2}\, \partial_\mu \widetilde \phi \Big) - B \,(\partial \cdot A) + \dfrac{B^2}{2}  \nonumber\\
&+&{\cal B} \Big(\dfrac{1}{2}\,\varepsilon_{\mu\nu\sigma} \,\partial^\mu B^{\nu\sigma} \Big) 
- \dfrac{{\cal B}^2} {2} + B^\mu \Big( \partial^\nu B_{\nu\mu} 
- \dfrac{1}{2}\, \partial_\mu \phi \Big) - \dfrac{1}{2}\, B^\mu B_\mu, \\
&&\nonumber\\
\label{19}
{\cal L}^{(2)}_{({\cal \bar B}, \bar B)} &=& \dfrac{1}{2}\, \bar {\cal B}^\mu \bar {\cal B}_\mu 
+ \bar {\cal B}^\mu \Big(\varepsilon_{\mu\nu\sigma} \partial^\nu A^\sigma 
+ \dfrac{1}{2}\, \partial_\mu \widetilde \phi \Big) - B \, (\partial \cdot A) + \dfrac{B^2}{2} \nonumber\\
&+& {\cal B} \Big(\dfrac{1}{2}\,\varepsilon_{\mu\nu\sigma} \,\partial^\mu B^{\nu\sigma} \Big) 
-\dfrac{{\cal B}^2} {2} - {\bar B}^\mu \Big( \partial^\nu B_{\nu\mu} 
+ \dfrac{1}{2}\, \partial_\mu \phi \Big) - \dfrac{1}{2}\, {\bar B}^\mu {\bar B}_\mu. 
\end{eqnarray}
where the subscripts on the above Lagrangian densities denote the {\it short} forms of the Nakanishi-Lautrup 
auxiliary fields that have been invoked to linearize the kinetic and gauge-fixing terms of the combined system 
of the free Abelian 2-form and 1-form gauge theories. To be more specific, we note that, in the Lagrangian density~\eqref{18}, 
a pair of auxiliary fields (${\cal B}_\mu, \, B $) have been invoked to linearize the kinetic and gauge-fixing terms of the
$3D$ Abelian 1-form gauge theory, respectively. On the other hand, another pair of auxiliary fields (${\cal  B}, \, B_\mu $) have been 
incorporated into the theory to linearize the kinetic and gauge-fixing terms of the $3D$ Abelian 2-form theory, respectively. 
In exactly similar fashion, we can specify the {\it concise} form of the subscript on the Lagrangian density~\eqref{19}
 where the replacements for the pairs of auxiliary fields  are: $({\cal B}_\mu, \, B ) \Rightarrow (\bar {\cal B}_\mu, \, B )$ and  
$({\cal  B}, \, B_\mu ) \Rightarrow ({\cal  B}, \, \bar B_\mu )$, respectively. In the above Lagrangian densities~\eqref{18} 
and \eqref{19}, an additional pair of dynamical fields ($\widetilde \phi, \; \phi $) have been added at appropriate places 
with proper mass dimensions in natural units (where $\hbar = c = 1$). These fields are nothing but the pseudo-scalar 
and scalar fields, respectively (cf. Sec.~\ref{sec5} for more discussions). The above linearizations are 
{\it general} in the sense that we have taken into
account {\it both} the signs that can be associated with the fields ($\widetilde \phi, \; \phi $). In other words, 
a close look at~\eqref{18} and~\eqref{19} demonstrates that both the Lagrangian densities can be {\it exchanged}
with each-other by taking into account the replacements: 
$ (\widetilde \phi, \; \phi ) \leftrightarrow (-\,\widetilde \phi, \; -\, \phi ), \;
({\cal B}_\mu, \, B ) \leftrightarrow (\bar {\cal B}_\mu, \, B ), \;
({\cal  B}, \, B_\mu ) \leftrightarrow ({\cal  B}, \, \bar B_\mu )$.  At this juncture, it is interesting to point 
out that the discrete duality symmetry transformations~\eqref{16},  existing for the gauge-fixed Lagrangian 
density~\eqref{12}, can be generalized into the following transformations:
\begin{eqnarray}\label{20}
&& A_\mu \longrightarrow \,\mp\, \dfrac{i}{2}\, \varepsilon_{\mu\nu\sigma}\, B^{\nu\sigma},  \quad 
B_{\mu\nu} \longrightarrow \,\pm i\,\varepsilon_{\mu\nu\sigma}\, A^\sigma, \quad 
B \longrightarrow \, \mp\, i\, {\cal B}, \qquad {\cal B} \longrightarrow \,\pm\, i\, B,\nonumber\\
&&B_\mu \longrightarrow \,\mp\, i \,{\cal B}_\mu, \qquad 
{\cal B}_\mu \longrightarrow\, \; \pm\, i\, B_\mu,  \qquad \bar B_\mu \longrightarrow \,\mp\, i\, \bar {\cal B}_\mu, 
\qquad  \bar {\cal B}_\mu \longrightarrow\, \pm\, i \,\bar B_\mu, \nonumber\\
&&  \phi \longrightarrow \, \mp\, i\, \widetilde \phi, \quad 
\widetilde \phi \longrightarrow \,\pm\, i \,\phi.
\end{eqnarray} 
The above discrete duality symmetry transformations are respected by the {\it linearized } 
versions of the Lagrangian densities~\eqref{18} and~\eqref{19}.

For the sake of brevity, in our present endeavor, we consider  {\it only} (i) the off-shell nilpotent 
(co-)BRST symmetry transformations which are the generalizations of the {\it classical} (dual-)gauge 
symmetry transformations~\eqref{14} and~\eqref{2} to the {\it quantum} level, and (ii) the (co-)BRST 
invariant Lagrangian density ${\cal L}_{B}$ which is the  generalization of the {\it classical} linearized version  of the
Lagrangian density ${\cal L}^{(1)}_{({\cal B}, B)} $ [cf. Eq.~\eqref{18}]  to its {\it quantum} level. 
In other words, we promote the Lagrangian density~\eqref{18} to incorporate into  itself  the Faddeev-Popov (FP) 
ghost terms corresponding to the $3D$ field-theoretic model of the  combined system of the  free Abelian 1-form 
and 2-form gauge theories as follows
\begin{eqnarray}\label{21}
{\cal L}_{B} &=& \dfrac{1}{2}\, {\cal B}^\mu {\cal B}_\mu 
- {\cal B}^\mu \Big(\varepsilon_{\mu\nu\sigma} \partial^\nu A^\sigma 
- \dfrac{1}{2}\, \partial_\mu \widetilde \phi \Big) - B \,(\partial \cdot A) + \dfrac{B^2}{2}  \nonumber\\
&+&{\cal B} \Big(\dfrac{1}{2}\,\varepsilon_{\mu\nu\sigma} \,\partial^\mu B^{\nu\sigma} \Big) 
- \dfrac{{\cal B}^2} {2} + B^\mu \Big( \partial^\nu B_{\nu\mu} - \dfrac{1}{2}\, \partial_\mu \phi \Big) - \dfrac{1}{2}\, B^\mu B_\mu \nonumber\\
&-& \big(\partial^\mu \bar C^\nu - \partial^\nu \bar C^\mu \big) \big(\partial_\mu C_\nu \big) 
- \dfrac{1}{2} \Big(\partial  \cdot C - \dfrac{\lambda}{4} \Big) \rho - \dfrac{1}{2} \Big(\partial  \cdot \bar C 
+ \dfrac{\rho}{4} \Big) \lambda \nonumber\\
& -& \dfrac{1}{2} \partial^\mu \bar \beta \, \partial_\mu \beta - \partial^\mu \bar C \, \partial_\mu C,
\end{eqnarray} 
where the FP-terms, corresponding to the free {\it massless} Abelian 2-form theory, have been 
systematically derived in our earlier works~\cite{RPM6,RPM9}. In the latter work~\cite{RPM9},  we
have considered the BRST-quantized version of the $4D$ St{\"u}ckelberg-modified {\it massive} Abelian 2-form 
theory and the FP-ghost term (i.e. $- \,\partial^\mu \bar C \, \partial_\mu C $) for the free Abelian 1-form 
theory is the standard book material. In the above Lagrangian density~\eqref{21}, the Lorentz vector fermionic 
(i.e. $C_\mu^2 = \bar C_\mu^2 = 0, \, C_\mu \, \bar C_\nu + \bar C_\nu \,  C_\mu = 0, \,C_\mu \, C_\nu + C_\nu \,  C_\mu = 0,$ etc.) 
(anti-)ghost fields $(\bar C_\mu)C_\mu$ are the generalizations of the gauge transformation parameter
$\Lambda_\mu$ [cf. Eq.~\eqref{2}] and they carry the ghost numbers $(-1)+1$, respectively, and the bosonic 
(anti-)ghost fields $(\bar \beta)\beta$ are endowed with the ghost numbers $(-2)+2$. Another
pair of the fermionic (i.e. $C^2 = \bar C^2 = 0, \, C \, \bar C + \bar C \, C = 0$) (anti-)ghost fields $(\bar C)C$ 
carry the ghost numbers $(-1)+1$ and they are the generalizations of the Lorentz scalar gauge  transformation 
 parameter $\Lambda$ [cf. Eq.~\eqref{2}] and the auxiliary (anti-)ghost fields $(\rho)\lambda$ are endowed
with ghost numbers $(-1)+1$, respectively. These (anti-)ghost fields are required to maintain the sacrosanct 
property of unitarity at any arbitrary order of perturbative computations. The other symbols in~\eqref{21} carry the 
{\it same} meaning as explained in the context of the Lagrangian density in~\eqref{18}.

The following infinitesimal, continuous and off-shell nilpotent (i.e. $s_d^2 = 0, \; s_b^2 = 0$) 
(co-)BRST transformations $(s_d)s_b$, namely;
\begin{eqnarray}\label{22}
&& s_d B_{\mu \nu} = \varepsilon_{\mu\nu\sigma}\, \partial^\sigma \bar C, \qquad \quad
s_d A_\mu = - \varepsilon_{\mu\nu\sigma}\,\partial^\nu \bar C^\sigma, 
\quad \qquad s_d \bar C_\mu  = -\, \partial_\mu \bar \beta,  \nonumber\\
&&s_d  C_\mu  = - \,{\cal B}_\mu,  \qquad\quad 
 s_d  C =  {\cal B}, \qquad \quad
s_d \beta = - \,\lambda,  \qquad \quad
s_d \widetilde \phi  = \rho, \nonumber\\
&& s_d \big[B_\mu, \, {\cal B}_\mu, \, B,\, {\cal B}, \phi, \bar \beta,\, \bar C,\, \rho,  \lambda  \big] = 0,  \qquad
\end{eqnarray} 
\begin{eqnarray}\label{23}
&& s_b B_{\mu \nu} = - \big(\partial_\mu C_\nu - \partial_\nu C_\mu\big),  \qquad \quad s_b A_\mu = \partial_\mu C, \qquad  
\quad s_b C_\mu  = -\, \partial_\mu \beta,  \nonumber\\ && s_b \bar C_\mu  = B_\mu, \qquad \quad 
 s_b \bar C = B, \qquad \quad 
s_b \bar \beta = -\, \rho, \qquad \quad  s_b \phi  = \lambda, \nonumber\\
&& s_b \big[B_\mu,\, {\cal B}_\mu, \, B,\, {\cal B}, \widetilde \phi, \beta,\, C,\, \rho, \, \lambda  \big] = 0, \qquad 
\end{eqnarray} 
are the {\it symmetry} transformations\footnote{These off-shell nilpotent symmetry transformations are {\it distinctly} 
different from the corresponding transformations in our very recent publications~\cite{RPM19,RPM20}.} 
of the action integral: $S = \int d^3 x\, {\cal L}_{B}$ because the Lagrangian density ${\cal L}_{B} $ 
[cf. Eq.~\eqref{21}] transforms to the total spacetime derivatives under the above nilpotent (co-)BRST 
[i.e. (dual-)BRST]  symmetry transformations as follows:
\begin{eqnarray}\label{24}
s_d {\cal L}_{B} = - \partial_\mu\Big[ \big(\partial^\mu \bar C^\nu - \partial^\nu \bar C^\mu \big)\, {\cal B}_\nu 
- {\cal B} \, \partial^\mu \bar C - \frac{1}{2}\, \rho\, {\cal B}^\mu - \frac{1}{2}\, \lambda\, \partial^\mu \bar \beta \Big],
\nonumber\\
s_b \, {\cal L}_{B} = -\partial_\mu \Big[ \big(\partial^\mu C^\nu - \partial^\nu C^\mu \big)\, B_\nu + B \, \partial ^\mu C 
+ \frac{1}{2}\,\lambda\, B^\mu - \dfrac{1}{2}\,\rho\, \partial^\mu \beta  \Big].
\end{eqnarray} 
In addition to the above  off-shell nilpotent (co-)BRST  symmetry transformations, the Lagrangian density~\eqref{21} 
respects a set of discrete duality symmetry transformations in the {\it ghost-sector} which is listed as follows:
\begin{eqnarray}\label{25}
&&C_\mu \,\longrightarrow\, \pm\,i\,\bar C_\mu, \qquad   \bar C_\mu \,\longrightarrow\, \pm\,i\, C_\mu, \qquad  
C \,\longrightarrow\, \mp\,i\,\bar C,  \qquad 
\bar C \,\longrightarrow\, \mp\,i\, C, \nonumber\\
&& \beta \,\longrightarrow\, \pm\,i\,\bar \beta, \qquad  \quad \bar \beta \,\longrightarrow\, \mp\,i\,\beta,  \qquad\;\;\;\;
\rho \,\longrightarrow\, \mp\,i\,\lambda, ~\qquad  \lambda \,\longrightarrow\, \mp\,i\,\rho. 
\end{eqnarray} 
Thus, it is clear that the total (co-)BRST invariant Lagrangian density \eqref{21} respects the discrete duality symmetry 
transformations in the {\it non-ghost} sector [cf. Eq.~\eqref{20}] as well as in the {\it ghost-sector} 
[cf. Eq.~\eqref{25}]. To sum-up, the Lagrangian density \eqref{21} respects the infinitesimal, continuous and 
nilpotent symmetries [cf. Eqs.~\eqref{22}, \eqref{23}] and a set of  discrete {\it duality} symmetry transformations. 
The {\it latter} is the sum of the discrete symmetry transformations that have been listed in~\eqref{20} and~\eqref{25}. 
We call these discrete symmetry transformations as the {\it duality} symmetry transformations because the basic 1-form 
and 2-form fields are connected with each-other [cf. Eq.~\eqref{20}] and they are {\it dual} to each-other on a flat $3D$ 
Minkowskian spacetime manifold [cf. footnote below Eq.~\eqref{16}].


\section{Bosonic Symmetry and Algebraic Structures}
\label{sec4}


It is straightforward to note, from the off-shell nilpotent symmetries in our equations~\eqref{22} and \eqref{23}, that the anticommutator 
(i.e. $\{ s_b, \; s_d \} $) between them is {\it not} equal to zero. In fact, this anticommutator defines a set of
the bosonic symmetry (i.e. $s_\omega = \{ s_b, \; s_d \} $) transformations, under which, the fields of \eqref{21} transform as:
\begin{eqnarray}\label{26}
&& s_\omega B_{\mu\nu} =  \varepsilon_{\mu\nu\sigma}\, \partial^\sigma B  
+ \big(\partial_\mu {\cal B}_\nu - \partial_\nu {\cal B}_\mu\big), \qquad \quad
s_\omega A_\mu = -\varepsilon_{\mu\nu\sigma}\, \partial^\nu B^\sigma + \partial_\mu {\cal B}, \nonumber\\
&& s_\omega C_\mu = \partial_\mu  \lambda, \quad 
s_\omega \bar C_\mu = \partial_\mu \rho, \quad
s_\omega \big[B_\mu, \, {\cal B}_\mu, \, B, \, {\cal B},\, \phi,\, \widetilde\phi,\, \beta,\, \bar \beta, \rho, \, 
\lambda, \, C, \,\bar C \big] = 0. 
\end{eqnarray} 
At this stage, it is worthwhile to mention that under the above bosonic symmetry transformations, the (anti-)ghost fields {\it either} do not 
transform at all {\it or} they transform up to the $U(1)$ gauge symmetry-type transformations. Furthermore, it is very interesting to point
out that, in their operator forms, the nilpotent (co-)BRST symmetry transformations and the bosonic symmetry transformation obey the
following algebra, namely;
\begin{eqnarray}\label{27}
 &&s_b^2 = 0, \quad \qquad s_d^2 = 0, \quad \qquad s_\omega = \big \{ s_b, \; s_d \big \} \equiv \big (s_b + s_d \big )^2, \nonumber\\
 && \big [s_\omega, \; s_b \big ] = 0, \quad \qquad 
\big [s_\omega, \; s_d \big ] = 0, \quad \qquad  \big \{ s_b, \; s_d \big \} \neq 0,
\end{eqnarray} 
which establish the fact that the bosonic symmetry transformation, in its operator form, commutes with {\it both} the 
off-shell nilpotent (co-)BRST symmetry transformation operators. The algebra~\eqref{27} is reminiscent of the 
following algebra obeyed by a set of {\it three}  de Rham cohomological operators of differential geometry~\cite{RPM14,RPM15,RPM16,RPM17,RPM18} 
\begin{eqnarray}\label{28}
&&d^2 = 0, \quad \qquad \delta^2 = 0, \quad \qquad \Delta = \big \{ d, \; \delta \big \} \equiv \big (d + \delta\big )^2 , \nonumber\\
&& \big [\Delta, \; d \big ] = 0, \quad \qquad \big [\Delta, \; \delta \big ] = 0, \quad \qquad  \big \{ d, \; \delta \big \} \neq 0,
\end{eqnarray}
where $d$ (with $d^2 = 0$) is the exterior derivative, $\delta = \pm\, *\, d\, *$ (with $\delta^2 = 0$) is the 
co-exterior (or dual-exterior) derivative and $\Delta = (d + \delta)^2$ is the Laplacian operator (which is always 
positive definite). Here the symbol $*$ stands for the Hodge duality operator on the compact spacetime 
manifold on which the cohomological operators are defined. The above algebraic structure, satisfied by the de Rham cohomological
operators, are popularly known as the Hodge algebra of differential geometry (see, e.g.~\cite{RPM14,RPM15,RPM16,RPM17,RPM18} for details).

We end this short section with a couple of useful remarks. First of all, at the algebraic level, it is very 
tempting to appreciate the similarity between the above algebraic relationships in~\eqref{27} and \eqref{28} 
which allow us to state that there is a  one-to-one mapping\footnote{We have obtained the one-to-one mapping 
because we have considered only the Lagrangian density~\eqref{21} which is the generalization of the linearized 
{\it classical} version of the  gauge-fixed Lagrangian density~\eqref{18} to its {\it quantum} level. If we had 
considered the {\it quantum} version of the linearized {\it classical} version of the gauge-fixed Lagrangian 
density~\eqref{19} {\it together} with the {\it above} generalization, we would have ended up with the two-to-one 
mapping between the symmetry transformation operators and the de Rham cohomological operators of differential 
geometry as we have obtained in our earlier works~\cite{RPM6,RPM7,RPM8,RPM9}.} between the symmetry transformation operators 
($s_b, \; s_d, \; s_\omega $) and the de Rham cohomological operators ($d, \; \delta, \; \Delta$). To corroborate
this claim, we have to recall our key observations in Secs.~\ref{sec2} and~\ref{sec3} where we have established that, under the 
(dual-)gauge and (co-)BRST symmetry transformations, the (i) gauge-fixing terms (owing their origin to the 
co-exterior derivative), and (ii) kinetic terms (owing their existence to the exterior derivative) for the 
{\it gauge} fields remain invariant, respectively. Hence, the above identification: 
$(s_b, \; s_d, \; s_\omega) \Leftrightarrow (d, \; \delta, \; \Delta)$ is realistic at the algebraic level. 
Second, the (co-)exterior derivatives $(\delta)d$, as pointed out earlier, are connected by the relationship: 
$\delta = \pm\, *\, d\, *$. Within the framework of BRST formalism, {\it this} relationship is realized in 
terms of the nilpotent (co-)BRST  transformation operators and the discrete  transformation
operators  [cf. Eqs.~\eqref{20}, \eqref{25}] as
\begin{eqnarray}\label{29}
 s_d \, \Phi = \pm\, *\, s_b \, *\, \Phi, \qquad \Phi = B_{\mu\nu}, A_\mu,  B_\mu, {\cal B}_\mu,  \bar C_\mu,  C_\mu,  \phi,  \widetilde \phi,
B,  {\cal B},  \bar \beta,  \beta, \bar C,  C, \rho, \lambda,  
\end{eqnarray}
where the symbol $*$ stands for the discrete duality symmetry transformations. Thus, we observe that the 
interplay between the continuous and discrete symmetry operators of our theory provides the physical realization 
of the relationship: $\delta = \pm\, *\, d\, *$. In the above equation~\eqref{29}, as is obvious, the generic 
field of the Lagrangian density~\eqref{21} has been denoted by the field $\Phi$. The ($\pm $) signs, 
on the r.h.s. of the above equation \eqref{29}, are dictated by a couple of successive operations 
of the discrete duality symmetry transformation operator  on the generic field $\Phi$ of the Lagrangian 
density~\eqref{21} as follows~\cite{RPM31}:
\begin{eqnarray}\label{30}
 * \; \big (*\; \Phi \big ) = \pm\, \Phi.
 \end{eqnarray}
It is an elementary exercise to note that, for all the bosonic fields of our (co-)BRST invariant Lagrangian 
density~\eqref{21}, we have the {\it plus} sign on the r.h.s. of~\eqref{30} and, hence, in the equation~\eqref{29}, too. 
For instance, it is straightforward to check that: $*\,(*\, B_{\mu\nu}) = +\, B_{\mu\nu} $. As a consequence, we observe that: 
$s_d \,B_{\mu\nu} = +\, *\, s_b\, *\, B_{\mu\nu}$ is true. On the other hand, the signs on the r.h.s. of~\eqref{30} and~\eqref{29} 
are {\it negative} when we take into account the {\it fermionic} fields of the (co-)BRST invariant Lagrangian density~\eqref{21}. 
For instance, we can readily check that: $*\,(*\, C_{\mu}) = -\, C_{\mu} $. As a consequence, we have the algebraic relationship 
between the off-shell nilpotent (co-)BRST symmetry transformation operators [cf. Eqs.~\eqref{22}, \eqref{23}] as: 
$s_d \,C_{\mu} = -\, *\, s_b\, *\, C_{\mu}$ where the symbol $*$ stands for the set of discrete duality symmetry transformation 
operators that have been listed in the equations~\eqref{20} and~\eqref{25} for our (co-)BRST invariant Lagrangian density~\eqref{21}.
To be precise, these discrete duality symmetry transformations [cf. Eqs.~\eqref{20},~\eqref{25}] are present  in the non-ghost and 
ghost-sectors of our  field-theoretic model of the {\it combined} system of the BRST-quantized version of the $3D$ 
free Abelian 1-from and 2-form gauge theories.


\section{Conclusions}
\label{sec5}


In our present investigation, we have taken into account the {\it combined} system of the free $3D$ Abelian 1-form 
and 2-form gauge theories {\it together} and shown the existence of (i) the nilpotent BRST symmetry transformations, 
(ii) the nilpotent co-BRST symmetry transformations, and (iii) a bosonic symmetry transformation which is constructed 
from the anticommutator of the BRST and co-BRST symmetry transformations. Under the BRST symmetry transformations, 
the Lagrangian densities of the free Abelian 1-form and 2-form gauge theories remain invariant, separately and independently. However,
for the validity of the co-BRST symmetry invariance, we find that (i) the kinetic term of the Abelian 2-form field, and 
(ii) the FP-ghost term of the Abelian 1-form theory transform in such a manner that they produce a total spacetime 
derivative\footnote{In exactly similar fashion, we note that (i) the kinetic term of the Abelian 1-form gauge field, 
and (ii) the {\it basic} FP-ghost term of the Abelian 2-form theory transform to a total spacetime derivative 
under the co-BRST symmetry transformations thereby leading to the invariance of the action integral.}. 
The algebra obeyed by these {\it symmetry} operators [cf. Eq.~\eqref{27}] is reminiscent of the algebra satisfied by the  
de Rham cohomological operators of differential geometry [cf. Eq.~\eqref{28}].

One of the highlights of our present investigation is the existence of the discrete duality symmetry transformations 
in our BRST-quantized theory which has its mathematical origin in the Hodge duality operator (on the $3D$ flat Minkowskian 
spacetime manifold) as far as the {\it basic} gauge fields of the Abelian 1-form and 2-form theories are concerned
[cf. Eq.~\eqref{16} and corresponding footnote]. The interplay between the discrete duality symmetry transformations 
[cf. Eqs.~\eqref{20}, \eqref{25}] and continuous symmetry transformations [cf. Eqs.~\eqref{22}, \eqref{23}] for the (co-)BRST 
invariant Lagrangian density~\eqref{21} are responsible for the physical realization [cf. Eq.~\eqref{29}]  of the well-known 
relationship (i.e. $\delta = \pm\, *\, d\, * $) between the (co-)exterior derivatives of differential  
geometry~\cite{RPM14,RPM15,RPM16,RPM17,RPM18}. The most {\it basic} relationship~\eqref{16} between 
the gauge fields of the Abelian 1-form and 2-form theories establishes clearly that there is a duality symmetry 
between the free Abelian 2-form and 1-form {\it basic} fields\footnote{We would like to mention that these duality 
symmetry transformations have {\it not} been mentioned in  our very recent work~\cite{RPM20} on the {\it combined} system 
of the free $3D$ Abelian 1-form and 2-form gauge theories. This is why we have {\it not} been able to obtain 
the analogue of~\eqref{29}.}  in three (i.e. $D = 3$) dimensions of spacetime when they are taken {\it together} 
for specific purpose. It is interesting to add that these discrete duality symmetry transformations 
[cf. Eqs.~\eqref{20}, \eqref{25}] are also responsible for establishing a {\it direct} connection between the off-shell  
nilpotent (co-)BRST transformation operators (cf.~\ref{secA} for details).

We would like to say a few words about the presence of the scalar and pseudo-scalar fields 
(i.e. $\phi,\, \widetilde \phi $) in our (co-)BRST invariant Lagrangian density~\eqref{21}. It is very 
interesting to observe that  (i) the scalar field appears in our theory with a positive kinetic term, 
and (ii) the pseudo-scalar field is endowed with a negative kinetic term (cf. Sec.~\ref{sec1} for more details). 
The {\it latter} field (having the negative kinetic term) is one of the possible candidates for the ``phantom'' 
fields which have become quite  popular in the context of the cyclic, bouncing and self-accelerated 
cosmological models of the Universe (see, e.g.~\cite{RPM21,RPM22,RPM23,RPM24,RPM25} and references therein). Such massless fields 
(having {\it only} the negative kinetic term) are also relevant in the context of being one of the passable
 candidates  for dark energy (see, e.g.~\cite{RPM26,RPM27} for details) which are supposed to be responsible for the 
current observations of the accelerated cosmic expansion~\cite{RPM32,RPM33,RPM34,RPM35} of the Universe.

As far as our present $3D$ field-theoretic system is concerned, the pseudo-scalar field appears in the generalization of 
the kinetic term [$-\,\frac{1}{4}\, (F^{\mu\nu})^2 = -\,\frac{1}{2}\, (\varepsilon_{\mu\nu\sigma}\, \partial^\nu A^\sigma)^2 $]
 of the Abelian 1-form theory when it is expressed in terms of the Hodge dual of the Abelian 2-form: 
$ *\, F^{(2)} = * \, \big [\frac{1}{2!}\, F_{\mu\nu}\, (dx^\mu \wedge dx^\nu) \big ] 
\equiv (\varepsilon_{\mu\nu\sigma}\, \partial^\nu A^\sigma) \, dx^\mu$ which turns out to be a pseudo 1-form because
the quantity ($\varepsilon_{\mu\nu\sigma}\, \partial^\nu A^\sigma $) is an axial-vector. Here we have the freedom 
to add/subtract [cf. Eqs.~\eqref{18}, \eqref{19}] a pseudo 1-form $\widetilde \phi^{(1)} = (\partial_\mu \widetilde \phi)\, dx^\mu$ 
that is constructed from the derivative on a pseudo-scalar  field $\widetilde \phi$. This massless pseudo-scalar  field 
(with the negative kinetic term) is very important in our theory because it plays a crucial role in the existence of 
(i) the discrete duality symmetry transformations [cf. Eq.~\eqref{20}], and 
(ii) the non-trivial dual-BRST (i.e. co-BRST) symmetry transformations [cf. Eq.~\eqref{22}]. A close look at the 
equations~\eqref{18}, \eqref{19}, \eqref{20}, \eqref{22} and \ref{secA} demonstrates  that we have the nilpotent 
co-BRST symmetry transformations in our  {\it combined} field-theoretic system of the free Abelian 2-form and 
1-form gauge theory because of the presence of the pseudo-scalar  field. In contrast, the pure scalar field remains 
{\it inert} as far as the co-BRST symmetry transformations are concerned [cf. Eq.~\eqref{22}].
We would like to re-emphasize that the existence of (i) the nilpotent co-BRST symmetry transformations, and 
(ii) the pseudo-scalar field, is crucial for our $3D$ theory to be a field-theoretic example for Hodge theory within the 
framework of BRST formalism.

In our present investigation, we have  (i) {\it not} computed the Noether conserved charges corresponding 
to the off-shell nilpotent (co-)BRST symmetry transformations [cf. Eqs.~\eqref{22}, \eqref{23}] as well as  
the bosonic symmetry transformations [cf. Eq.~\eqref{26}], and (ii) {\it not} considered the generalization of the
{\it classical} gauge-fixed Lagrangian density~\eqref{19} to its {\it quantum} counterpart within the framework 
of BRST formalism. In this context, it is gratifying to state that, in our very recent paper [36], we have 
considered the {\it quantum} generalizations of  {\it both} the gauge-fixed Lagrangian densities~\eqref{18} and~\eqref{19} 
 and obtained the two-to-one mappings between the full set of symmetry transformation operators of the BRST-quantized $3D$ 
theory and the set of cohomological operators of differential geometry. In our future endeavors, we plan to devote 
time to address the above issues and compute the extended BRST algebra among the {\it appropriate} set of conserved 
charges of our $3D$ theory and establish its deep connection with the algebra obeyed by the de Rham cohomological 
operators of differential geometry~\cite{RPM14,RPM15,RPM16,RPM17,RPM18} so that we can prove that our present odd dimensional (i.e. $3D$)
field-theoretic model is a {\it perfect} example for Hodge theory. This proof will be different 
 from our earlier works~\cite{RPM6,RPM7,RPM8,RPM9} that are in the {\it even} dimensions of spacetime.

\section*{Acknowledgment}

One of us (RPM) would like to thank E. Harikumar, A. K. Rao , S. K. Panja,  Bhagya R. and H. Sreekumar  for fruitful
discussions on the main topic of our present endeavor.\\

\vskip 0.5cm

\noindent
{\bf Data Availability}\\

\noindent
No data were used to support this study.\\

\vskip 0.5cm

\noindent
{\bf Conflicts of Interest}\\

\noindent
The authors declare that there are no conflicts of interest.\\

\appendix

\section{On Direct Connection: Nilpotent $s_d$ and $s_b$}
\label{secA}
\renewcommand{\theequation}{A.\arabic{equation}}
\setcounter{equation}{0}
Besides the algebraic relationship we have obtained between the off-shell nilpotent ($s_d^2 = 0, \; s_b^2 = 0$) 
(co-)BRST symmetry transformation operators $(s_d)s_b$ in equation~\eqref{29}, there is  a {\it direct} relationship 
between {\it these} fermionic (i.e. off-shell nilpotent) symmetry transformation operators due to the presence of the discrete 
duality symmetry transformation operators [cf. Eqs.~\eqref{20}, \eqref{25}]. To corroborate this statement, let us pick-up 
the nilpotent BRST symmetry transformation: $s_b\, B_{\mu\nu} = - \,(\partial_\mu C_\nu - \partial_\nu C_\mu)$ and apply 
the discrete duality symmetry transformation operators~\eqref{20} and~\eqref{25}  on it as follows:
\begin{eqnarray}\label{A.1}
\big(* s_b \big) \big(* B_{\mu\nu} \big) = - * \big(\partial_\mu C_\nu - \partial_\nu C_\mu \big).
\end{eqnarray}
Using the discrete transformations~\eqref{20} and~\eqref{25} and taking into account the operation: $*\,s_b = s_d$, 
we obtain the following explicit relationship
\begin{eqnarray}\label{A.2}
s_d \Big(\pm \,i\, \varepsilon_{\mu\nu\sigma} \, A^\sigma \Big) 
= - \,\Big[\partial_\mu \big(\pm\, i \, \bar C_\nu \big) - \partial_\nu \big(\pm\,i\, \bar C_\mu \big) \Big],
\end{eqnarray}
which, finally, leads to the following 
off-shell nilpotent (i.e. $s_d^2 = 0 $) co-BRST symmetry transformation for the Abelian 1-form vector field $A_\mu$ as:
\begin{eqnarray}\label{A.3}
s_d\, A_\mu = -\, \varepsilon_{\mu\nu\sigma}\, \partial^\nu \bar C^\sigma.
\end{eqnarray}
Thus, starting from the BRST symmetry transformation for an antisymmetry tensor gauge field $B_{\mu\nu}$, 
we have obtained the co-BRST symmetry transformation for the Abelian 1-form vector field $A_\mu$ {\it correctly} 
due to the direct application of the discrete duality symmetry transformation operators [cf. Eqs.~\eqref{20}, \eqref{25}]
and taking into account the discrete duality symmetry transformation for the BRST symmetry transformation 
operator $s_b$ as: $*\, s_b = s_d$ (which leads to the derivation of the co-BRST symmetry 
transformation for $A_\mu$ field).

Against the backdrop of our above discussions, let us derive the nilpotent BRST symmetry transformation: 
$s_b\, B_{\mu\nu} = -\, (\partial_\mu C_\nu - \partial_\nu C_\mu)$ from the nilpotent co-BRST symmetry 
transformation: $s_d\, A_\mu = - \,\varepsilon_{\mu\nu\sigma}\, \partial^\nu \bar C^\sigma$. 
Once again, we exploit theoretical strength of the discrete duality symmetry transformations 
[cf. Eqs.~\eqref{20}, \eqref{25}] as follows
\begin{eqnarray}\label{A.4}
\big(* \,s_d \big) \big(* A_\mu \big) = -\, \varepsilon_{\mu\nu\sigma}\, \partial^\nu \big(*\, \bar C^\sigma \big),
\end{eqnarray}
where we take: $*\, s_d = -\, s_b$ as an input. We observe the following from~\eqref{A.4}, namely;
\begin{eqnarray}\label{A.5}
-\, s_b \Big(\mp \,\frac{i}{2}\, \varepsilon_{\mu\nu\sigma} \, B^{\nu\sigma} \Big) 
= -\, \varepsilon_{\mu\nu\sigma} \partial^\nu \big(\pm \,i\, C^\sigma \big).
\end{eqnarray}\label{A.6}
Simplifying the above relation by exploiting the straightforward property of the $3D$ 
Levi-Civita tensor,  we finally obtain the  nilpotent (i.e. $s_b^2 = 0 $) BRST transformation ($s_b $): 
\begin{eqnarray}
s_b\, B_{\mu\nu} = -\, (\partial_\mu C_\nu - \partial_\nu C_\mu).
\end{eqnarray}
Thus, we conclude that the  off-shell nilpotent ($s_d^2 = 0, \; s_b^2 = 0$) (co-)BRST symmetry 
transformation operators $(s_d)s_b$ are {\it directly} connected if we use the duality property: 
$*\, s_b = s_d,\; *\, s_d = - \,s_b$ and the discrete duality symmetry transformation operators 
[cf. Eqs.~\eqref{20}, \eqref{25}] judiciously. Ultimately, we have been able to establish that the mapping: 
$s_b \Leftrightarrow s_d$ can be realized (if we use {\it directly} the discrete duality symmetry 
transformation operators [cf. Eqs.~\eqref{20}, \eqref{25}] along with the inputs:  
 $*\, s_b = s_d$,\; $* \,s_d = -\, s_b$). We have been able to establish these kinds of relationships 
among {\it all} the  (co-)BRST symmetry transformation operators $(s_d)s_b$ [cf. Eqs.~\eqref{22}, \eqref{23}] as follows:
\begin{eqnarray}\label{A.7}
&& s_b \, A_\mu \; \Leftrightarrow \; s_d\, B_{\mu\nu}, \qquad \qquad
s_b\, B_{\mu\nu} \; \Leftrightarrow \; s_d \,A_\mu, \qquad \qquad
s_b\,  C_\mu  \; \Leftrightarrow \; s_d \,\bar C_\mu,  \nonumber\\
&& s_b\, \bar C_\mu  \; \Leftrightarrow \; s_d \,C_\mu,  \qquad
s_b\, \bar \beta \; \Leftrightarrow \; s_d \,\beta, \qquad 
s_b\, \phi \; \Leftrightarrow \; s_d \,\widetilde \phi, \qquad 
s_b\, \bar C \; \Leftrightarrow \; s_d \,C,\nonumber\\
&& s_b \big[B_\mu,\, {\cal B}_\mu,\, B,\, {\cal B},\, \widetilde \phi, \, \beta, \, C, \, \rho, \, \lambda  \big] \; \Leftrightarrow \;
s_d \big[{\cal B}_\mu,\, B_\mu,\, {\cal B},\, B, \, \phi, \, \bar \beta, \, \bar C, \, \lambda, \, \rho  \big].
\end{eqnarray}
It is very interesting to point out that the nilpotent ($s_b^2 = 0, \; s_d^2 = 0$) BRST ($s_b$) and co-BRST ($s_d$) 
symmetry transformation operators are connected with each-other (i.e $s_b \;\Leftrightarrow \; s_d$) through the fields 
that are intimately related to each-other by the direct duality symmetry transformations [cf. Eqs.~\eqref{20},\eqref{25}]
in our $3D$ field-theoretic model.

We end this Appendix with the following remarks. First, we note that the {\it basic} discrete duality symmetry transformations 
[cf. Eq.~\eqref{16}] exist between the Abelian 1-form field $A_\mu$ and 2-form field $B_{\mu\nu}$ at the level of {\it classically} 
gauge-fixed Lagrangian density [cf. Eq.~\eqref{12}]. Second, these duality relationships are promoted at the quantum level when 
the Faddeev-Popov (FP) ghost terms are added to the gauge-fixed Lagrangian density where we have to take the total discrete duality 
symmetry transformations for the non-ghost fields [cf. Eq.~\eqref{20}] as well as for the (anti-)ghost fields [cf. Eq.~\eqref{25}]. 
Third, we have noted in this Appendix that the total set of discrete duality symmetry transformations [cf. Eqs.~\eqref{20},\eqref{25}] 
provide a direct mapping between the off-shell nilpotent BRST and co-BRST symmetry transformation operators [cf. Eq.~\eqref{A.7}] 
provided we use: $*\, s_b = s_d$,\; $* \,s_d = -\, s_b$. Finally, it is very interesting to point out that the duality symmetry 
transformations between the  BRST and co-BRST symmetry operators (i.e.  $*\, s_b = s_d$,\; $* \,s_d = -\, s_b$)  are just like the 
electromagnetic duality symmetry transformations between the electric and magnetic fields of the source-free Maxwell's theory of 
electromagnetism in vacuum.

\end{document}